Tidal Evolution of a Secularly Interacting Planetary System

By Richard Greenberg
and
Christa Van Laerhoven

Lunar and Planetary Laboratory
University of Arizona
1629 East University Boulevard
Tucson, AZ 85721-0092

Short title: Secular evolution with tides

Corresponding author:
Richard Greenberg
greenberg@lpl.arizona.edu

Tidal Evolution of a Secularly Interacting Planetary System

**Abstract**

In a multi-planet system, a gradual change in one planet's semi-major axis will affect the eccentricities of all the planets, as angular momentum is distributed via secular interactions. If tidal dissipation in the planet is the cause of the change in semi-major axis, it also damps that planet's eccentricity, which in turn also contributes to the evolution of all the eccentricities. Formulae quantifying the combined effects on the whole system due to semi-major axis changes, as well as eccentricity damping, are derived here for a two-planet system. The CoRoT 7 system is considered as an example.

**Key Words**: Celestial mechanics, Planets and satellites: dynamical evolution and stability, planetary systems, planet-star interactions

**1. Introduction**

In a multi-planet system, damping of the orbital eccentricity of one planet, by tides for example, affects the eccentricities of all the orbits through secular interactions and a consequent exchange of angular momentum (e.g. Wu and Goldreich 2002). Given enough time, all the orbits would become circularized. Classical secular theory (e.g. Brouwer and Clemence 1961, Murray and Dermott 1999, Barnes and Greenberg 2006) provides the basis for an analytically tractable approach to this behavior. Even incorporating a process that tends to damp an eccentricity, the differential equations that describe the behavior remain linear, assuming none of the eccentricities is too large, and thus the equations are readily solvable (e.g. Chiang and Murray 2002, Wu and Goldreich 2002, Zhang and Hamilton 2003). Where the eccentricities are large, other approaches are necessary, such that of Mardling (2007). However, wherever it is applicable the classical approach does have the advantage of offering complementary insight into the process. The solution reduces to a set of eigenmodes, each of which is damped at a characteristic rate. Thus it readily explains why some planetary systems tend to have the orientations of their major axes (i.e. the lines of apsides) aligned before having all the eccentricities damp away on a much longer timescale. Even where eccentricities are too large for the classical theory to apply with precision, the character of the behavior found by Mardling (2007) is qualitatively consistent with the predictions of classical secular theory.

The effects of eccentricity damping may be important in many extra-solar systems where the innermost planet is close enough to undergo tidal damping of its eccentricity (Jeffreys 1961, Goldreich and Soter 1966). However, tides also change the semi-major axis of the planet (Goldreich and Soter 1966), which must be considered in order to account properly for how the tidal effects are transmitted to the other planets (Jackson et al. 2008). (Throughout this paper "changing the semi-major axis" refers to a change in the length of the semi-major axis $a$; We use "line of apsides" when refering to the orientation of this axis.) The importance of the role of a changing semi-major axis in modifying secular evolution was noted by Wu and Goldreich (2002), who evaluated it under certain special assumed conditions. Here we derive a more general solution for the effect of a slowly varying semi-major axis on the eccentricities of all the planets in a system.

Wu and Goldreich considered a two-planet system in which the amplitude of one of the eigenvectors is zero, i.e. one of the two eigenmodes has already died out. They also assumed for purposes of their analysis that the eccentricities were small enough that the secular theory could be developed to low order in the eccentricities. The system that motivated their study, and to which they applied the theory, was that of HD 83443. The alignment of lines of apsides in the initially reported elements for that system (Mayor et al. 2000) helped justify the assumption that one eigenmode had died out, although the reported eccentricity of the outer planet (0.42) made the small-e assumption questionable for that system. That issue became mute when the second planet was later found to be non-existent (Butler et al. 2002). Nevertheless, the analysis by Wu and Goldreich may become relevant again as new systems continue to be discovered.

However, in order to be more widely applicable, expressions are needed for the effect of a changing semi-major axis on the damping of eccentricities that avoid the assumption that only one eigenmode has a non-zero amplitude. In fact, such equations are essential if we are to consider tidal evolution, going back in time, through conditions where all eigenmodes are significant (i.e. before any have damped away). Such a theory is developed here. Like Wu and Goldreich, we assume that the eccentricities are small enough for a low-order analysis, and we consider a coplanar two-planet system. Our general approach could readily be extended to a system with any number of planets. It can also accommodate any process that gradually changes one (or more) of the semi-major axes, as well as damping one or more of the eccentricities. We show that the resulting equations for tidal evolution reduce to those of Wu and Goldreich in the special case that they considered.

The next section (2) briefly reviews the standard secular theory, largely to define the notation used here. Section III then shows the derivation of the damping rates for both eigenfrequencies, due to the changes in semi-major axis $a_1$ and eccentricity $e_1$ of the inner planet. Section IV shows that these results reduce to those of Wu and Goldreich in the special case that they considered, and the conclusions are summarized in Section 6.

## 2. Secular theory and notation

Secular theory has been described in detail in several textbooks and numerous published papers, with various sets of notation. Here we adopt the notation of Wu and Goldreich in order to facilitate comparison and continuity with their development. Tides tend to reduce the eccentricity of the inner planet according to

$$de_1/dt = -e_1/\tau \tag{1}$$

where $\tau$ describes the rate of damping due to any cause. Note that in the astrophysical literature $\tau$ is often called the tidal damping "timescale". However, the damping is not exponential, because $\tau$ is a function of $e_1$ itself as well as of $a_1$, which also varies due to tides. Thus the actual damping timescale, even for a single planet system, may be quite different from $\tau$ (Jackson et al. 2008). Wu and Goldreich considered the damping due to tidal dissipation within the planet itself and defined $\tau$ accordingly in terms of planetary physical properties for a particular tidal model. However, in a planetary system, other processes may come into play, such as tides raised by the planet on the star. So we can be more general by not specifying the definition of $\tau$ any further than what is shown in Eq. (1). Also, whereas the tide considered by Wu and Goldreich conserves the orbital angular momentum of the system (such that $da_1/dt \propto de_1/dt$), our theoretical approach will accommodate any process that slowly changes the semi-major axis of the inner planet even if angular momentum is not conserved.

The eccentricities $e$ and longitudes of pericenter $\varpi$ can be expressed, as is traditional, in terms of the elements $h$ and $k$ defined as:

$$h = e \sin \varpi \quad \text{and} \quad k = e \cos \varpi. \tag{2}$$

Lagrange's equations for the variation of the elements, retaining only secular terms to lowest order in $e$ yield (e.g. Brouwer and Clemence 1961, Murray and Dermott 1999):

$$dh_1/dt = A_{11}k_1 + A_{12}k_2 - h_1/\tau \tag{3a}$$

$$dk_1/dt = -A_{11}h_1 - A_{12}h_2 - k_1/\tau \tag{3b}$$

$$dh_2/dt = A_{21}k_1 + A_{22}k_2 \tag{3c}$$

$dk_2/dt = -A_{21}h_1 - A_{22}h_2$            (3d)

where the coefficients $A$ are each a function of the masses of the planets and stars and of the semi-major axes, and the last terms in (3a) and (3b) follow from Eq. (1). As defined here both $A_{11}$ and $A_{22}$ are positive; $A_{12}$ and $A_{21}$ are negative.

Solving this set of linear differential equations we find two eigenfrequencies:

$$g = \frac{A_{11} + A_{22} + (1/\tau)i \pm \sqrt{(A_{11} - A_{22} + (1/\tau)i)^2 + 4A_{12}A_{21}}}{2} \quad (4)$$

Following Wu and Goldreich, we define $g_p$ and $g_m$ as the values of $g$ with the plus or minus sign, respectively, in front of the radical. More generally, throughout this paper, any symbols with a $p$ or an $m$ subscript refer to portions of the solution associated with those frequencies. For tides, the value of $1/\tau$ is small, so we can linearize Eq. (4) to obtain:

$$2g = A_{11} + A_{22} \pm S + [1 \pm (A_{11} - A_{22})/S](1/\tau)i \quad (5)$$

where

$$S \equiv \sqrt{(A_{11} - A_{22})^2 + 4A_{12}A_{21}} \quad (6)$$

For each solution ($g_p$ or $g_m$), the values of the $h$'s and $k$'s oscillate sinusoidally with a frequency given by the real part of (5), while the amplitudes of the oscillations gradually damp down at a rate given by the imaginary part of (5). The character of the solution has been described in detail elsewhere many times. Here we recall a few salient features, useful in the derivation that follows in Section 3. As illustrated in Fig. 1, in an ($h,k$) coordinate system the vector ($h_{1p},k_{1p}$) has a nearly constant magnitude, $e_{1p}$, and its direction β rotates with an angular velocity $g_p$. The vector ($h_{2p},k_{2p}$) also has a nearly constant magnitude, $e_{2p}$, and its direction is exactly opposite that of vector ($h_{1p},k_{1p}$). Similarly, the vectors ($h_{1m},k_{1m}$) and ($h_{2m},k_{2m}$) have fixed magnitudes, $e_{1m}$ and $e_{2m}$ respectively, and rotate with angular velocity $g_m$. For this eigenfrequency, the two vectors point in the same direction given by angle $\alpha$.

The solution to Eqs. (3) also gives the following ratios (equivalent to the eigenvectors):

$$e_{1p}/e_{2p} = -(A_{11} - A_{22} + S)/2A_{21} \qquad (7a)$$

$$e_{1m}/e_{2m} = (A_{11} - A_{22} - S)/2A_{21} \qquad (7b)$$

Note that the right sides of Eqs. (7) are positive. Here, following the convention of Wu and Goldreich, we define all e components as positive, bearing in mind that the actual eigenvector has the two components in the p mode in opposite directions as shown in Fig. 1.

Considering the imaginary part of g (from Eq. 5), we have the damping rates:

$$\dot{e}_{1p}/e_{1p} = \dot{e}_{2p}/e_{2p} = -[1 + (A_{11} - A_{22})/S]/(2\tau) \qquad (8a)$$

$$\dot{e}_{1m}/e_{1m} = \dot{e}_{2m}/e_{2m} = -[1 - (A_{11} - A_{22})/S]/(2\tau) \qquad (8b)$$

Note that the initial conditions (the values of $e_1$, $e_2$, $\varpi_1$ and $\varpi_2$) combined with Eqs. (7) yield the values of $e_{1p}$, $e_{2p}$, $e_{1m}$, $e_{2m}$, $\alpha$, and $\beta$. (Here $\alpha$ and $\beta$ are the phases of the m and p solutions, respectively, as shown in Fig. 1.)  Now suppose that there is a small instantaneous change in $a_1$ (called $\delta a_1$), but in no other of the keplerian orbital elements. The values of all the coefficients $A$ change slightly because they are functions of $a_1$. As a result, the values of $e_{1p}$, $e_{2p}$, $e_{1m}$, $e_{2m}$, $\alpha$, and $\beta$ would change. Thus, while Eqs. (8) describe changes due to damping of $e_1$, there are additional changes due to any small or slow change in $a_1$. These changes are evaluated in the next section.

## 3. Effect of changing semi-major axis

According to the geometry shown in Fig. 1,

$$e_1 \cos \varpi_1 = e_{1m} \cos \alpha + e_{1p} \cos \beta \qquad (9a)$$

$$e_1 \sin \varpi_1 = e_{1m} \sin \alpha + e_{1p} \sin \beta \tag{9b}$$

$$e_2 \cos \varpi_2 = e_{2m} \cos \alpha - e_{2p} \cos \beta \tag{9c}$$

$$e_2 \sin \varpi_2 = e_{2m} \sin \alpha - e_{2p} \sin \beta \tag{9d}$$

If there is a small change in $a_1$, the constants of integration may change, but the elements on the left sides of Eqs. 9 do not change. Differentiating these equations yields

$$0 = \delta e_{1m} \cos \alpha - e_{1m} \sin \alpha \, \delta\alpha + \delta e_{1p} \cos \beta - e_{1p} \sin \beta \, \delta\beta \tag{10a}$$

$$0 = \delta e_{1m} \sin \alpha + e_{1m} \cos \alpha \, \delta\alpha + \delta e_{1p} \sin \beta + e_{1p} \cos \beta \, \delta\beta \tag{10b}$$

$$0 = \delta e_{2m} \cos \alpha - e_{2m} \sin \alpha \, \delta\alpha - \delta e_{2p} \cos \beta + e_{2p} \sin \beta \, \delta\beta \tag{10c}$$

$$0 = \delta e_{2m} \sin \alpha + e_{2m} \cos \alpha \, \delta\alpha - \delta e_{2p} \sin \beta - e_{2p} \cos \beta \, \delta\beta \tag{10d}$$

Eqs. (10) provide four equations for the six unknown variations ($\delta e_{1p}$, $\delta e_{2p}$, $\delta e_{1m}$, $\delta e_{2m}$, $\delta\alpha$, and $\delta\beta$). In addition, we have the known ratios given by the two Eqs. (7). These six equations allow us to solve for the variations.

First, we can eliminate $\delta\alpha$ and $\delta\beta$ from Eqs. (10), leaving us (after some algebraic and trigonometric manipulation) with the two equations:

$$(e_{2m} \delta e_{1m} - e_{1m} \delta e_{2m}) \cos \theta + (e_{2m} \delta e_{1p} + e_{1m} \delta e_{2p}) = 0 \tag{11a}$$

$$(e_{2p} \delta e_{1p} - e_{1p} \delta e_{2p}) \cos \theta + (e_{2p} \delta e_{1m} + e_{1p} \delta e_{2m}) = 0 \tag{11a}$$

where $\theta = \beta - \alpha$. Next we define functions $F_p$ and $F_m$ by rewriting Eqs. (7) as

$$e_{1p} = F_p(a_1) \, e_{2p} \tag{12a}$$

$$e_{1m} = F_m(a_1) \, e_{2m} \tag{12b}$$

(In the notation of Wu and Goldreich, the right hand side of either of those equations was represented by a function $f(a_1, e_2)$. Remember they assumed that either $e_{1p} = e_{2p} = 0$ or $e_{1m} = e_{2m} = 0$.) From Eqs. (12a & b) and (7), we also have the useful relationship:

$$F_p(a_1) + F_m(a_1) = - S/A_{21} > 0 \tag{12c}$$

Differentiating Eqs. (12a & b) yields

$$\delta e_{1p} = F'_p(a_1) \, \delta a_1 \, e_{2p} + F_p(a_1) \, \delta e_{2p} \tag{13a}$$

$$\delta e_{1m} = F'_m(a_1) \, \delta a_1 \, e_{2m} + F_m(a_1) \, \delta e_{2m} \tag{13a}$$

where the prime (′) indicates a first derivative with respect to $a_1$. With Eqs. (11) and (13) we have four equations with four unknowns.

Solving these equations yields

$$\delta e_{2p} = - [(e_{2m} \, F'_m \cos \theta + e_{2p} \, F_p) / (F_p + F_m)] \, \delta a_1 \tag{14a}$$

$$\delta e_{2m} = - [(e_{2p} \, F'_p \cos \theta + e_{2m} \, F_m) / (F_p + F_m)] \, \delta a_1 \tag{14b}$$

with $\delta e_{1p}$ and $\delta e_{1m}$ then given by (13).

From the definitions of $F_p$ and $F_m$ (compare Eqs. (12) and (7)), we have $F_p + F_m = S/A_{21}$, so we can write a simpler form of (14):

$$\delta e_{2p} = - (A_{21}/S)(e_{2m} \, F'_m \cos \theta + e_{2p} \, F_p) \, \delta a_1 \tag{15a}$$

$$\delta e_{2m} = - (A_{21}/S)(e_{2p} \, F'_p \cos \theta + e_{2m} \, F_m) \, \delta a_1 \tag{15b}$$

Now, if $a_1$ varied at a known rate, we could simply replace $\delta a_1$ with $da_1/dt$ to obtain expressions for the rate of change of each component of the eccentricities. This procedure gives rates that are functions of $\theta$, but as long as $a_1$ varies slowly, the rates can be averaged over each cycle of $\theta$, of course taking into account any dependence of $da_1/dt$ on $\theta$. Those rates would be in addition to the effect of eccentricity damping given by Eq. (8). This general analytic approach could be applied to determine how secular interactions distribute among any number of planets the effects of any physical process (e.g. tides) that acts on the semimajor axis and/or eccentricity of any one (or more) of them.

Next we consider the specific process addressed by Wu and Goldreich: tidal dissipation within the inner planet, which affects both $e_1$ and $a_1$, such that angular momentum is conserved. For any process that conserves angular momentum, with $de_1/dt$ given by (1),

$$da_1/dt = -2e_1^2 a_1/\tau. \tag{16}$$

We know that $e_1$ generally varies considerably over each cycle of $\theta$ as shown in Fig. 1. The law of cosines yields

$$e_1^2 = e_{1m}^2 + e_{1p}^2 + 2\, e_{1m} e_{1p} \cos\theta \tag{17}$$

If we substitute (17) for $e_1^2$ in (16) and then put (16) into (15) in place of $\delta a_1$, we obtain the instantaneous rate of change of $e_{2p}$ and of $e_{2m}$. More useful is to average these rates over a cycle of $\theta$, which yields

$$de_{2p}/dt = -(2a_1 A_{21}/S)[F'_m e_{1m} e_{1p} e_{2m} + F'_p (e_{1m}^2 + e_{1p}^2)\, e_{2p}]/\tau \tag{18a}$$

$$de_{2m}/dt = -(2a_1 A_{21}/S)[e_{1m} e_{1p} e_{2p} F'_p + F'_m (e_{1m}^2 + e_{1p}^2)\, e_{2m}]/\tau \tag{18b}$$

Then, similarly inserting (16) and (17) into (13) and averaging over $\theta$ yields

$$de_{1p}/dt = -(2a_1/\tau)\,\{(A_{21}/S)F'_m F_p e_{1m} e_{1p} e_{2m} + F'_p (1+F_p A_{21}/S)(e_{1m}^2+e_{1p}^2)\, e_{2p}\} \tag{18c}$$

$$de_{1m}/dt = -(2a_1/\tau)\,\{(A_{21}/S)F'_p F_m e_{1m} e_{1p} e_{2p} + F'_m (1+F_m A_{21}/S)(e_{1m}^2+e_{1p}^2)\, e_{2m}\} \tag{18d}$$

Adding the rates given in (8) to those in (18) gives the complete set of expressions for the damping of all four eccentricity components:

$$de_{1p}/dt = (A_{21}/S)\, F_p\, e_{1p}/\tau - (2a_1 A_{21}/S)[F'_m F_p e_{1m} e_{1p} e_{2m} - F'_p F_m (e_{1m}^2+e_{1p}^2)\, e_{2p}]/\tau \tag{19a}$$

$$de_{2p}/dt = (A_{21}/S)\, F_p\, e_{2p}/\tau - (2a_1 A_{21}/S)[F'_m e_{1m} e_{1p} e_{2m} + F'_p (e_{1m}^2 + e_{1p}^2)\, e_{2p}]/\tau \tag{19b}$$

$$de_{1m}/dt = (A_{21}/S)\, F_m\, e_{1m}/\tau - (2a_1 A_{21}/S)[F'_p F_m e_{1m} e_{1p} e_{2p} - F'_m F_p (e_{1m}^2+e_{1p}^2)\, e_{2m}]/\tau \tag{19c}$$

$$de_{2m}/dt = (A_{21}/S)\, F_m\, e_{2m}/\tau - (2a_1 A_{21}/S)[F'_p e_{1m} e_{1p} e_{2p} + F'_m (e_{1m}^2 + e_{1p}^2)\, e_{2m}]/\tau \tag{19d}$$

The above expressions have been algebraically simplified by using the definitions of $F_m$ and $F_p$. Note that on the right side of each of the above four equations, the first term represents the effect of tidal damping of $e_1$ and the remainder of the expression is the result of tidal variation of $a_1$.

## 4. A special case: A single eigenfrequency

In the special case where either $e_{1m} = e_{2m} = 0$ or $e_{1p} = e_{1p} = 0$, it can be shown that the more general results derived in the previous section agree with those of Wu and Goldreich. For example consider their expression for the damping of $e_{2p}$ where they assumed $e_{1m} = e_{2m} = 0$:

$$de_{2p}/dt = -(e_{2p}/\tau)(1 - 2e_{2p}e_{2p}F'_p a_1) / [1 + (J_2/J_1)(e_{2p}/e_{1p})^2] \qquad (20)$$

where $J_1$ and $J_1$ are the angular momenta of the orbits, and we have converted their $f$ function to our notation. It can readily be shown that, to lowest order in eccentricities, $J_2/J_1 = A_{21}/A_{21}$. Also, the ratio $e_{2p}/e_{1p}$ is given by Eq. (7a). With these substitutions and considerable algebraic rearrangement, Eq. (19) becomes:

$$de_{2p}/dt = -[1 + (A_{21} - A_{21})/S]/(2\tau) - 2a_1(A_{21}/S)[F'_p e_{1p}^2 e_{2p}]/\tau \qquad (21)$$

The first term in Eq. 21 is identical to the damping rate given in Eq. (8a), which is the system's response to the tidal effect on the inner planet's eccentricity. The second term is identical to the rate given by Eq. (18a) with $e_{1m}$ and $e_{2m}$ set to 0. In other words, Eq. (21) is the same as our result (19b) for this case. Thus our result is identical to the result of Wu and Goldreich for the special case that they considered.

We can also use our result to check Wu and Goldreich's implicit assumption that if one eigenfrequency has zero amplitude, it will remain zero going back or forward in time so that their solution remains valid. For example, if $e_{1m} = e_{2m} = 0$, then $de_{1m}/dt$ and $de_{2m}/dt$ must both be identically zero for the Wu and Goldreich result to be meaningful. Inspection of our solution shows this to be the case. Thus, Wu and Goldreich's equations are valid at least for the special case in which one eigenfrequency has zero amplitude. However, it could not be applied under any other circumstances. In other words, they would not be applicable to any real system unless one of the two eigenfrequencies had already effectively died away. But, in such a case, the Wu and Goldreich solution could not be used in any study that attempted to go back in time to reconstruct the history of the orbital evolution, because going back in time the amplitudes of both eigenmodes would be expected to grow. The application to the putative system HD 83443 probably was not valid for that reason, although the issue is moot because the system was later shown not to exist, at least not in the form that had been reported earlier.

## 5. A numerical example: CoRoT 7

The method of analysis derived here will likely be widely applicable as the anticipated large number of discoveries of new planetary systems proceeds over the coming years. Of course the approach will only apply to systems for which the orbital eccentricities are small enough to justify application of classical second-order secular theory and for which processes (like tides) are likely to act directly on the *e* and *a* values of at least one planet. Even if such systems are a small fraction of the systems to be discovered, there are likely to many of them.

While the greatest usefulness of this approach is still to come in the near future, we can use parameters from systems already known to assess whether the new tidal damping terms developed here are likely to be large enough to be worth consideration as new systems are discovered. The CoRoT 7 system is a useful example. Only two planets have been confirmed, so the equations derived in Section 3 are relevant without further development. The eccentricities are probably small, in fact too small to have been detected so far. And the inner planet is so close to the star that past (and perhaps current) tidal evolution is likely. In fact the outer planet is also very close to the star, but for this example we will assume that it is not directly affected by tides, except through the secular interactions with the inner planet. (The derivation shown in Sections 2 and 3 could be extended by the same method to a system of any number of planets, in which any number are directly affected by tides or other dissipative processes. However, for this example we are considering only the case derived in Section 3, with two planets and the inner one undergoing tidal dissipation.)

For this example we use the most recent orbital fit for CoRoT 7 by Ferraz-Mello et al. (2011). While there remains considerable uncertainty regarding the parameters of this system, for our illustrative purposes the exact values are not critical. We thus adopt semi-major axis values of 0.0175 AU and 0.0456 AU, with planetary masses of 8 and 15 Earth masses, respectively. In this solution no non-zero eccentricity values are detected, so we assume the values are small enough to justify the classical secular theory.

For this system, the eigenfrequencies (Eq. 5) are 0.675°/yr and 0.134°/yr for eigenmodes p and m, respectively. First, consider the effect of the tide on the eccentricity $e_1$ of the inner planet, ignoring for the moment the effect on $a_1$. With the eccentricity damped according to Eq. (1), the damping rates (from Eq. 8) for the two eigenmodes are -0.955/$\tau$ and -0.045/$\tau$, respectively. With this solution, mode p would damp away relatively quickly, leaving mode m to damp slowly, over a much longer time. During this time, the lines of apsides of the two planets would be aligned, and the ratio of the eccentricities e1/e2 would be fixed at a value of 0.38, a ratio given by the eigenvector. This "quasi-fixed-point" behavior would endure until mode m eventually damps away, a typical solution as discussed in the introduction (Section 1).

Now, if we take into account the tidal variation in semi-major axis $a_1$ that accompanies the tidal effect on $e_1$ (*i.e.* the new analysis developed in this paper), the evolution is quite different. According to Eq. 19:

$$de_{1p}/dt/e_{1p} = -0.955\,(1 - 20.5 e_{2p}^2)/\tau \tag{22a}$$

$$de_{2p}/dt/e_{2p} = -0.955\,(1 + 430 e_{2p}^2)/\tau \tag{22b}$$

$$de_{1m}/dt/e_{1m} = -0.045\,(1 + 9500 e_{2p}^2 + 20.5 e_{2m}^2)/\tau \tag{22c}$$

$$de_{2m}/dt/e_{2m} = -0.045\,(1)/\tau \tag{22d}$$

Here terms within the parentheses with very small coefficients have been ignored. Within the parentheses in each equation, the first term (1) represents the part due to damping of $e_1$, while the other terms result from the tidal effect on $a_1$.

For our illustrative purposes, suppose $e_{1p} \sim e_{2m} \sim 0.1$. In this case, the eigenvectors (Eq. 7) give $e_{1m} = 0.038$ and $e_{2p} = 0.012$. With these values, even as the eccentricities undergo their periodic secular variations, neither ever gets much larger than ~0.1.

Evaluating Eqs. 22 with these values inserted shows that for the faster-damping mode p, $de_{2p}/dt/e_{2p}$ is about 7% faster than the inner planet's $de_{1p}/dt/e_{1p}$, a significant difference from the usual e-damping solution in which these quantities are equal. What is more, for mode m the damping rate $de_{1m}/dt/e_{1m}$ is nearly three times greater than for the outer planet. Even as mode p damps away, the much longer-lived eigenmode m will retain most of its amplitude. Then, according to Eqs. 22c and 22d, $de_{1m}/dt/e_{1m}$ is 20% faster than $de_{2m}/dt/e_{2m}$. Thus, rather than being in a quasi-fixed-point condition with the ratio of $e_1/e_2$ fixed, the inner planet's eccentricity damps considerably faster than the other planet's. This result is thus very different from the usual quasi-fixed-point solution in which the ratio of the eccentricities remains constant. Thus we see that it is crucial to take into account the role of changing semi-major axis values in driving damping of the eigenvector components of the eccentricities.

**6. Discussion**

We have derived formulae for the damping of all four of the eigenvector components that describe the evolution of eccentricities in a two-planet system. This solution does not require the assumption that the components corresponding to one of the eigenmodes are zero. Our analysis still does require that the eccentricities be small enough for classical secular theory to apply. Nevertheless, the formulae that we have derived may be used to explore the past and future tidal evolution of observed systems with multiple planets, including some that are close enough to experience tidal dissipation.

The specific formulae derived here apply to a two-planet system, but the derivation could readily be extended by the same method to any number of planets. It could also be applied to account for tidal dissipation in more planets than just the innermost one, a real possibility as numerous new systems continue to be discovered.

Here we have considered in detail a case where the dominant process driving orbital change is tidal dissipation within the innermost planet, or any process that similarly conserves angular momentum. However, the general method of our derivation could readily be applied to any other processes that tend gradually to alter eccentricities and semi-major axes. Such processes might include tides raised on the star by planets, gas drag, or continual interaction with a ring of small particles. As the number of known planetary systems and knowledge of their properties continues to grow, the analytical approach and formulae developed here may be useful for reconstructing the orbital histories of those systems and thus constraining the processes of planetary formation and evolution.

**Acknowledgments**

This paper was improved considerably thanks to the comments of an anonymous referee, which among other things included the suggestion to include the discussion in Section 5. Comments by Douglas Hamilton were also crucial. We thank Fred Rasio for his personal attention to this paper. This work was supported by a grant from NASA's Planetary Geology and Geophysics program, while that program still supported research in this area.

**Figure Caption**

Figure 1: The components of the eccentricities. For the "m" eigenmode, the eccentricity components $e_{1m}$ and $e_{2m}$ are aligned, with the rotating orientation $\alpha$. For the "p" eigenmode, the eccentricity components $e_{1p}$ and $e_{2p}$ are anti-aligned, with the rotating orientation $\beta$. The eccentricities $e_1$ and $e_2$ are the vector sums as shown. An instantaneous, infinitesimal change in a semi-major axis would preserve $e_1$ and $e_2$, but change the various components slightly. Thus subsequent behavior of $e_1$ and $e_2$ would be modified.

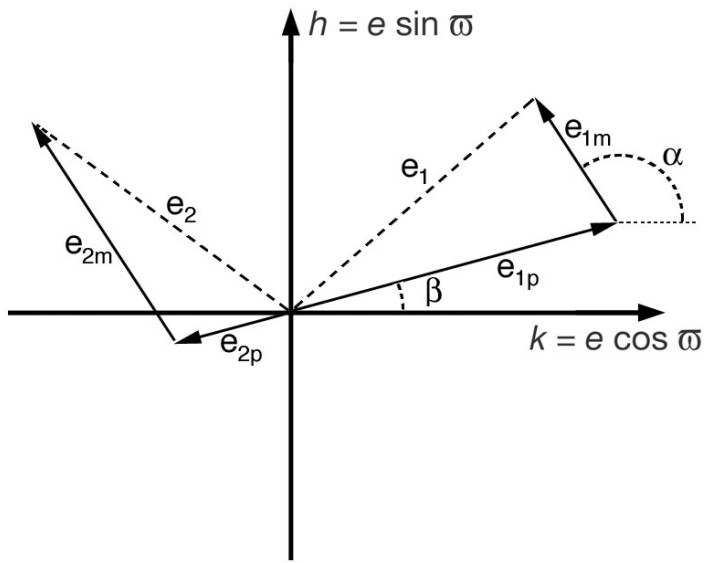